\documentclass[namedreferences]{solarphysics}
\usepackage[optionalrh]{spr-sola-addons} % For Solar Physics
\usepackage{graphicx,color}        % For eps figures, newer & more powerfull
\usepackage{color}           % For color text: \color command
\usepackage{url}             % For breaking URLs easily trough lines
\newcommand{\arcsec}{^{\prime\prime}}

\begin{document}

\begin{article}

\begin{opening}

\title{{Two Step Filament Eruption During 14--15 March 2015}}

\author[addressref={aff1},corref,email={rchandra.ntl@gmail.com}]{\inits{R.}\fnm{R.}~\lnm{Chandra}}%\sep
\author[addressref=aff2]{\inits{B.}\fnm{B.}~\lnm{Filippov}}%\sep
\author[addressref=aff1]{\inits{R.}\fnm{R.}~\lnm{Joshi}}%\sep
\author[addressref=aff3]{\inits{B.}\fnm{B.}~\lnm{Schmieder}}%\sep
%\author{\inits{}\fnm{}~\lnm{}\orcid{}}
%\author{P.~\surname{Author-a}$^{1}$\sep
%        E.~\surname{Author-b}$^{1}$\sep
%        M.~\surname{Author-c}$^{2}$
%       }

%   \institute{$^{1}$ First affiliation
%                     email: \url{e.mail-a} email: \url{e.mail-b}\\
%              $^{2}$ Second affiliation
%                     email: \url{e.mail-c} \\
%             }
\address[id=aff1]{Department of Physics, DSB Campus, Kumaun University, Nainital -- 263001, India}
\address[id=aff2]{Pushkov Institute of Terrestrial Magnetism, Ionesphere and Radio Wave Propagation, Russian Academy of Sciences, Troitsk, Moscow, Russia}
\address[id=aff3]{Observatoire de Paris, LESIA, UMR8109 (CNRS), F-92195, Meudon Principal
Cedex, France}

%\author{R. ~\surname{Chandra}\sep B. {Filippov}\sep R.~\surname{Joshi}\sep
%        B.~\surname{Schmieder}\sep
%       }
%
\runningauthor{R. Chandra {\it et al.}}
\runningtitle{Two Step Filament Eruption During 14--15 March 2015}
%\institute{R.~\surname{Chandra}\sep R.~\surname{Joshi}\\
%Department of Physics, DSB Campus, Kumaun University, Nainital -- 263001, India\\
%email: \url{rchandra.ntl@gmail.com}, \url{reetikajoshi.ntl@gmail.com}}
%
%\institute{B.~\surname{Filippov}\\
%Pushkov Institute of Terrestrial Magnetism, Ionesphere and Radio Wave Propagation, Russian Academy of Sciences, Troitsk, Moscow, Russia
%email: \url{bfilip@izmiran.ru} }
%
%\institute{B.~\surname{Schmieder}\\
%Observatoire de Paris, LESIA, UMR8109 (CNRS), F-92195, Meudon Principal
%Cedex, France\\
%email: \url{brigitte.schmieder@obspm.fr}}
%

\begin{abstract}
We present here an interesting two--step filament eruption during 14--15 March 2015. The filament was located in NOAA AR 
12297 and associated with a halo Coronal Mass Ejection (CME). We use observations from the 
{\em Atmospheric Imaging Assembly} (AIA) and {\em Heliospheric Magnetic Imager} (HMI) instruments 
onboard the {\em Solar 
Dynamics Observatory} (SDO), and from the {\em Solar and Heliospheric Observatory} (SOHO) {\em Large Angle and 
Spectrometric Coronagraph} (LASCO). We also use H$\alpha$ data from the {\em Global Oscillation Network Group} (GONG) telescope
and the {\em Kanzelhoehe Solar Observatory}. The filament shows a first step eruption on 14 March 2015 and it stops 
its rise at a projected altitude $\approx$ 125 Mm on the solar disk.
It remains at this height for $\approx$ 12 hrs. Finally it  eruptes on 15 March 2015 and produced a halo CME. 
We also find jet activity in the active region during both days, which could help the filament 
de-stabilization and eruption.
The decay index is calculated to understand this two--step eruption. 
The eruption could be due to the presence of successive instability--stability--instability zones as the filament is rising.

\end{abstract}

\keywords{Sun - filament: Sun - corona: Sun-jets: Sun - magnetic}
\end{opening}
%%%%%%%%%%%%%%%%%%%%%%%%%%%%%%%%%%%%%%%%%%%%%%%%%%%%%%%%%%%%%%%%%%%%%%%%%

\section{Introduction}
     \label{S-Intro}

Solar filaments (known as prominences, when they appear on the limb) are dense and cool material 
suspended in the hot solar corona along polarity inversion lines (PILs). They are found in magnetic dip regions. 
The filaments can be located in quiet as well as in solar active regions. The existence, stability and disappearance of 
filaments are nicely described in recent reviews \citep{Mackay10,Labrosse10,Parenti14}.  
There are two main magnetic configurations for the filaments, namely a 
sheared arcade and  a flux rope. In the sheared arcade configuration, the arcade connects the opposite polarities on 
either sides of a PIL, whereas in the case of the flux rope configuration the magnetic field has helical magnetic 
structure. 
Solar filament eruptions are also responsible for Coronal Mass Ejections (CMEs) and hence for  geomagnetic activity. 
Therefore the study of their stability and eruption is very important for space weather prediction.

Active region filaments (for example \opencite{Chandra10}) are usually associated more with solar flares in 
comparison to quiet region filaments \citep{Gosain16}. 
It is not necessary that every filament activation results in a CME. The result of the  activation of filaments can be 
a failed eruption \citep{Ji03,Torok05,David06,Kuridze13,Kushwaha15}, a partial eruption \citep{Gibson06,Liu14,Kliem14} or a full eruption \citep{Chandra16}. 
When a filament erupts partially or fully and moves away from the solar 
surface it becomes a CME. Therefore there is a spectrum from fully confined to 
fully eruptive filament activations. In the latest studies the {\em Solar Terrestrial Relations Observatory} (STEREO)
 has provided two view-points of filament eruptions. 
Combining the STEREO data with SDO data, three view-point studies 
have been done \citep{Filippov13,Filippov16a}.

A flux rope is in equilibrium 
due to the balance between the upward magnetic hoop--force and the downwards magnetic tension of the 
overlying coronal magnetic field. If due to some process like reconnection between new and existing 
magnetic flux the stability of the flux rope is lost, it starts to erupt. 
Several models have been proposed for  solar eruptions (for reviews see 
\opencite{Aulanier14,Filippov15,Schmieder15,Vrsnak14}). 
Catastrophic loss of equilibrium or the torus instability are the 
important mechanisms for solar eruptions \citep{Forbes91,Kliem06,Demoulin10}.
In these models, it is assumed that the overlying magnetic field (B$_{\textnormal {external}}$) decreases with the increase of the 
height ({\it h}) 
from the photosphere {\it i.e.} B$_{\textnormal {external}}$ $\propto$  {\it h}$^{-n}$. The flux rope becomes unstable when the 
decay index {\it n}  at its location becomes less than 
a critical value. According to simulations this value ranges from 1.3 to 1.75 
\citep{Torok05,Isenberg07,Aulanier10,Zuccarello17}. Using different observations including high resolution SDO and 
multi--view STEREO observations it was found that  this value lies between 1 and 1.5 in observed eruptions
\citep{Filippov01,Filippov13,Zuccarello14,McCauley15}.  

In addition to failed, partial and full filament eruptions, recently two filament eruptions were observed with a 
quasi-equilibrium state in the middle part of a two-step process \citep{Byrne14,Gosain16}. In these cases the filament 
starts to erupt and after attaining
some height it decelerates, stops and seems to be stable for some time. In case of \inlinecite{Byrne14} this time was 
rather short {\it i.e.} $\approx$ one hour and for the case of 
\inlinecite{Gosain16} the time was 15 hrs.  Such cases are very crucial and play an important role in 
understanding the space weather implications of filament eruptions. In this study, we analyze an active-region-filament eruption on 
14--15 March 2015, where we notice that 
after the initial eruption the filament stays for $\approx$ 12 hrs in the quasi-equilibrium state. 

The paper is organized as follows:
Section 2 presents the observational data, and the morphology of the filament eruption.
The time--distance analysis of the filament eruption is presented in Section 3. 
The associated CME is described in Section 4. To interpret the quasi--equilibrium of the filament, we present the
decay index calculation in Section 5.
Finally in Section 6, we discuss our results and conclude. 

%%%%%%%%%%%%%%%%%%%%%%%%%%%%%%%%%%%%%%%%%%%%%%%%%%%%%%%%%%%%%%%%%%%%%%%%%%%%%%%%

%%%%%%%%%%%%%%%%%%%%%%%%%%%%%%%%%%%%%%%%%%%%%%%%%%%%%%%%%%%%%%%%%%%%%%%%%%%%%%%%

\section{Observations}
     \label{obs}

The active region, NOAA AR 12297 produced one of the largest geomagnetic storm
 (D$_{\textnormal{st}}$ index $\approx$ $-223$ nT) of 
Solar Cycle 24
on 17 March 2015. This geomagnetic storm was associated with a filament eruption and a  {\em Geostationary Operational Environmental Satellite} 
(GOES) C9.1 class flare of 2015 
March 15. 
The active region NOAA AR 12297 was located at S22W25 on 15 March 2015 on the solar disk. 
Further this filament eruption was associated with a halo CME.
The filament eruption on 2015 March 15 was studied by \inlinecite{Wang16},  whileas the geomagnetic storm 
from this active region was investigated by \inlinecite{Liu15}. 
The eruption also produced a magnetic cloud (MC) studied by \inlinecite{cho17}.
They calculated the helicity sign in the MC and found 
it was associated with a positive helicity.
Our aim in this study is to investigate the filament
activation and eruption
of 14--15 March 2015. This filament was disturbed on 14 March 2015 and finally on 15 March 2015, it totally erupted. 
The event was observed by the {\em Atmospheric Imaging Assembly} (AIA, \opencite{Lemen12}) onboard the {\em Solar Dynamics Observatory (SDO, 
\opencite{Pesnell12})} in different UV and EUV wavelengths. 
The pixel size and cadence of the data is 0.6 arcsec and 12 s respectively. 
The event was also observed by the {\em Global Oscillation 
Network Group} (GONG) in the H$\alpha$ line center. The temporal and pixel 
resolution of GONG H$\alpha$ data are one min and one arcsec respectively.
For the magnetic field, we use the
data from {\em Helioseismic and Magnetic Imager} (HMI, \opencite{Scherrer12}) onboard SDO. 
The cadence of the HMI data is 45 s and the pixel resolution is 0.5 arcsec. 

The observational description of the filament activation and eruption on 14--15 March 2015 observed by different 
instruments is presented in the following sections.
%%%%%%%%%%%%%%%%%%%%%%%%%%%%%%%%%%%%%%%%%%%%%%%%%%%%%%%%%%%%%%%%

\subsection{SDO and H$\alpha$ Observations}
     \label{sdo}

\begin{figure}
\includegraphics[width=1.0\textwidth, clip=]{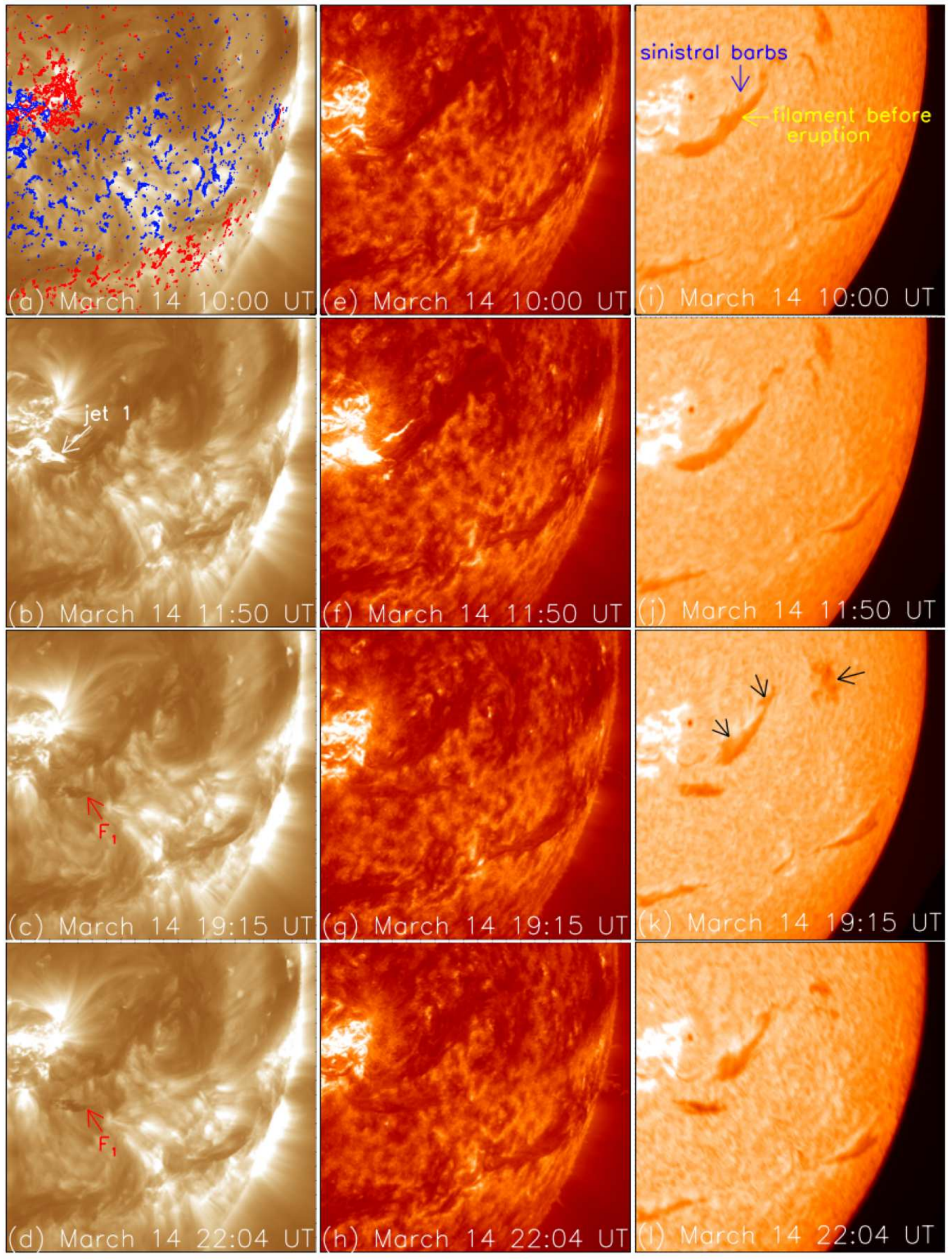}
\caption{Evolution of the filament on 14 March 2015. Left column : AIA 193 \AA; middle 
column: AIA 304 \AA; and right column: GONG H$\alpha$ observations. 
In panel (a) the AIA image is overlaid by HMI contours (levels: $\pm$ 50, $\pm$ 100, $\pm$ 200 G). Red and blue contours represents the positive and negative polarities respectively. 
The position of jet 1 is shown by the 
white arrow. The interrupted  eruption of the left part of the filament is shown 
by F$_1$ and by red arrows. The major remaining part of the filament is shown by black arrows. A sinistral barb is shown by a blue arrow.}   
\label{fig:morpho1}
\end{figure}

Figure \ref{fig:morpho1} shows the evolution of the filament on 14 March 2015 in AIA 193 \AA\ (first column), 304 \AA\ 
(second column), and GONG H$\alpha$ (third column). 
Figure \ref{fig:morpho1}a is overplotted with HMI magnetic field contours.
Before any activity in the 
filament it has a long sigmoidal shape (see Figure \ref{fig:morpho1}a, e, f, and also the yellow arrow).
 We can see the sinistral barbs in GONG H$\alpha$ images. 
According to \inlinecite{Martin98}, if the barbs in the filament are directed leftwards (rightwards) their
chiralities are sinistral (dextral) respectively. 
The sinistral (dextral) chiralities indicate positive (negative) twist/helicity
respectively. In our case the sinistral barbs are represented by blue arrow in Figure \ref{fig:morpho1}i.
This suggests the filament has a positive twist and hence  positive helicity. 
Since the filament is located in the southern hemisphere, this is in conformity with
the hemispheric rule of helicity. According to the helicity
hemispheric rule the majority of positive/negative helicity solar features located in the southern/northern hemisphere 
respectively \citep{Pevtsov95}.

Around 11:50 UT a jet activity started in the active 
region, which we name as jet 1 and mark by the white arrow in Figure \ref{fig:morpho1}b. This jet activity was visible in almost all EUV/UV 
and H$\alpha$ wavelengths. Due to this jet activity the left lower part of the sigmoidal filament was disturbed. 
Afterwards a strand of the filament separated from the main body, rose upwards and became stable at the projected altitude  $\approx$ 125 Mm.  
The broken part  of the sigmoidal filament is labeled by F$_1$ and indicated 
by the red arrow in Figure \ref{fig:morpho1}c, d. At the 
same time, we observe flare brightening in the active region which
was reported by GOES as a C2.6 class flare. 
The upper--right part of the filament channel, which was not disturbed in 
this period is shown by black arrows in Figure \ref{fig:morpho1}k.
LASCO C2 at $\approx$ 13:30 UT on March 14 also observed a partial 
halo CME. Later on this CME interacted 
with the CME of March 15 \citep{Liu15}. The CME on March 14 was associated with the small filament in the active region, which is different than the filament studied here.
A small filament eruption was associated with this CME as reported by \inlinecite{Wang16} 
and is named F$_4$ in their Figure 1a.   

\begin{figure}
\includegraphics[width=1\textwidth, clip=]{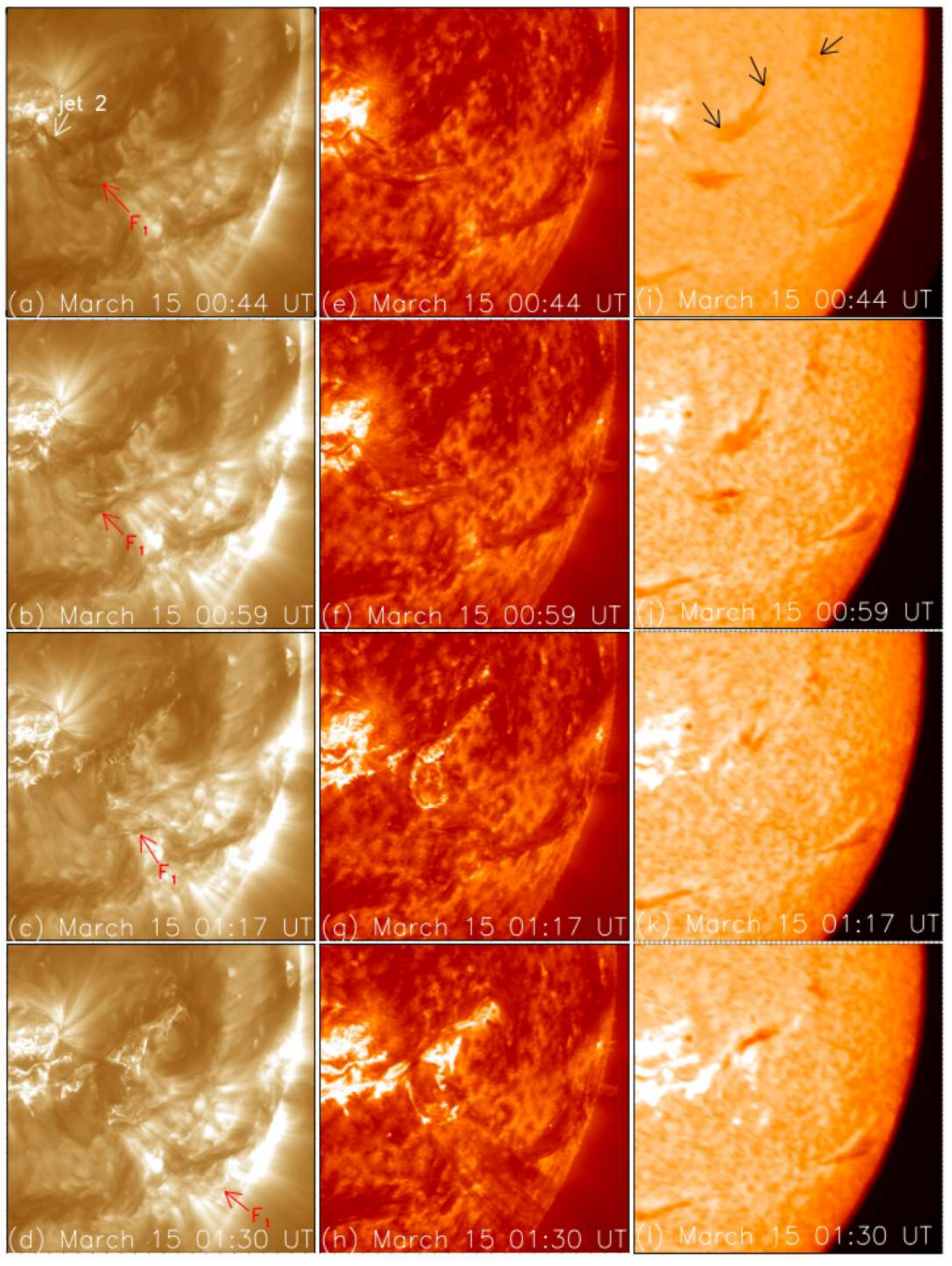}
\caption{Evolution of the filament on 15 March 2015. Left column : AIA 193 \AA; middle column: AIA 304 \AA; and right column: GONG 
H$\alpha$ observations. The position of jet 2 is shown by the white arrow. The erupting filament F$_1$  is shown by
 red arrows. The black arrows represent the failed erupted part of the filament.}   
\label{fig:morpho2}
\end{figure}

An interesting point is that the uplifted broken part F$_1$ from the main sigmoidal filament was stable in that 
particular location for a period of $\approx$ 12 hrs {\it i.e} 
up to 00:45 UT on 15 March 2015. Around 00:45 UT on 15 March 2015, we again 
observe the jet/surge activity in the active region towards the west-south direction which interacts with the filament F$_1$ as
 well as with the big northern filament (indicated by three black arrows). 
We label this jet as jet 2. After 00:45 UT the filament F$_1$ starts to erupt 
and finally it leaves the solar surface. 
The evolution of the erupting filament is shown in Figure \ref{fig:morpho2}. The left, middle and right column of 
Figure \ref{fig:morpho2} show the development of the erupting filament F$_1$ in AIA 193 \AA, 304 \AA, and H$\alpha$ 
wavelengths respectively. The eruption is associated with a 
GOES 9.1 class flare. 
The flare has a two--ribbon structure visible in different EUV and H$\alpha$ images. We can also see the ribbon separation 
as proposed by the CSHKP model \citep{Carmichael64,Sturrock66,Hirayama74,Kopp76}.
The flare was also observed in hard X-rays by the {\em Reuven Ramaty High Energy Solar Spectroscopic Imager} (RHESSI) 
satellite and studied by \inlinecite{Wang16}.

The major upper part of the sigmoidal filament channel (indicated by black arrows in Figure \ref{fig:morpho2}) 
is also perturbed during the ejection of jet 2. The material of the
filament rises up and later on it comes back down to the footpoints of the filaments. 
In AIA channels it becomes brighter and we can also see its twisted structure. 
From Figure \ref{fig:morpho2}d (see also movie in AIA 193 \AA), one can infer that 
the twist is right--handed, which is consistent with the sinistral filament 
chirality seen in H$\alpha$ data. Therefore, we see here the same sign of twist in the chromosphere, the upper
solar atmosphere and in the associated magnetic cloud. It seems that jet 2 injects impulse and heat into the filament at its eastern end.
Just after the maximum brightness of the mini-flare in the active region at 00:39 UT, dark and 
bright features appear in AIA 193 \AA\ images, which move from the flare place along the filament 
channel (see movie in AIA 193 \AA). Obviously the flare happens near the eastern end of the filament and 
jet 2 propagates along field lines belonging to the flux rope containing the filament. Accelerated and 
heated plasma of the jet influence significantly the equilibrium of the whole filament.
Previously static filament material starts to move along the helical field-lines of the flux rope. 
Thus the upper parts of the helices become visible as dark and bright (due to heating) threads. 
We observe intense field-aligned motions within the activated flux rope but it does not change 
significantly its position, and after the energetic phase of the event the filament is restored to its 
approximate initial state. We consider arguments for such behavior in Section 5.
During the activation the filament becomes less visible in the H$\alpha$ wavelength (see the H$\alpha$ 
images in the third column of Figure \ref{fig:morpho2}) most likely for two reasons. 
The Doppler shift in the moving material can move its H$\alpha$ line out of the filter passband 
(dynamic disappearance), and the heating of the filament can suppress the absorption of H$\alpha$ radiation (thermal disappearance).

\section{Time--Distance Analysis}
     \label{td}
\begin{figure}
\includegraphics[width=1.0\textwidth, clip=]{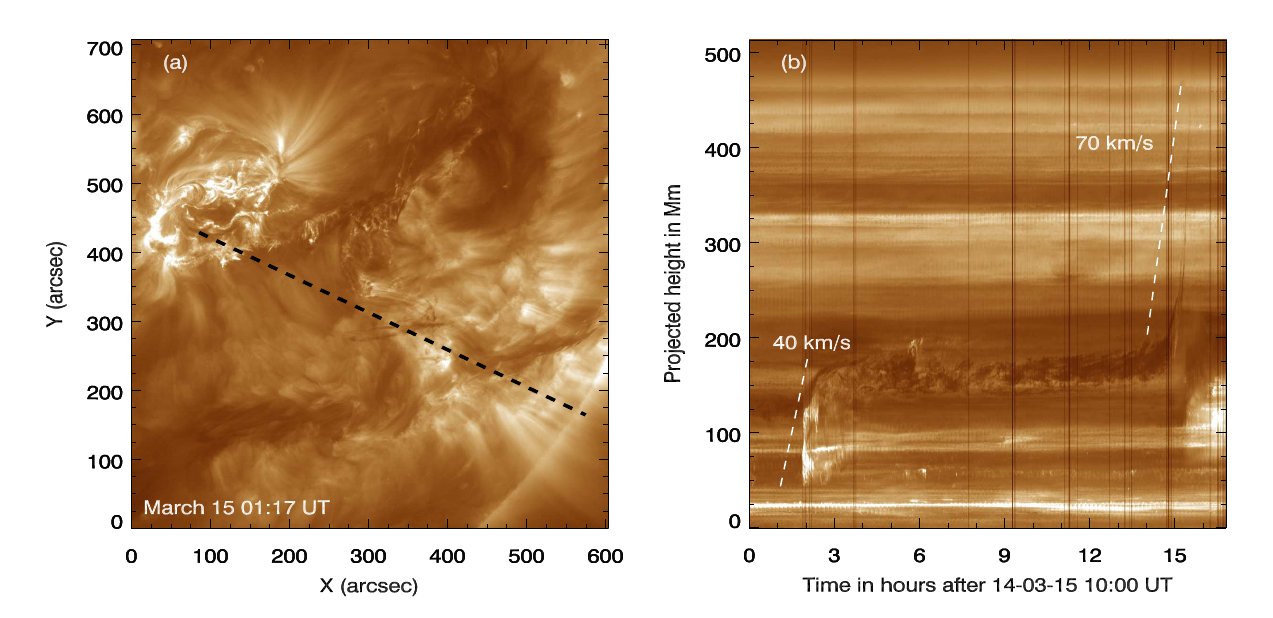}
\caption{(a) AIA 193 \AA\ image showing the location of a slit selected for the time--slice analysis. 
(b)Time-slice image of the eruption during 14--15 March 2015.}
\label{fig:slice}
\end{figure}

To investigate the height--time evolution of the filament eruption during 14--15 March 2015, we 
 select a slit along the
direction of the filament eruption in the AIA 193 \AA\ data. The position of the slit is shown by the 
dashed black line in Figure \ref{fig:slice}a.
The time-distance diagram is given in Figure \ref{fig:slice}b. In the time-distance diagram, we can clearly
see the two step eruption of the filament. 
In the first step, the filament starts to 
rise at $\approx$ 12:00 UT on 14 March 2015 and shifts to a projected altitude of 125 
Mm on the disk. 
Unfortunately we do not have STEREO observations, which
could tell us the true height of the eruption. We compute the speed of this eruption
and find it to be $\approx$ 40 km s$^{-1}$. After 13:00 UT the filament stops its rise and stays at the same 
height up to 00:45 UT on 15 March 2015. In the second step after 00:45 UT on 15 March 2015, the filament starts 
to rise and finally fully erupts. The calculated speed of the second step eruption is 70 km s$^{-1}$.

\section{LASCO Observations}
     \label{lasco}

The CME associated with the filament eruption on 15 March 2015 was observed with the LASCO instrument \citep{Bruecknet95}.
As reported in \inlinecite{Wang16}, during the C2.6 class flare of 14 March 2015 a
small filament erupted (see their Figure 1a, F$_4$ filament) and there was a slow CME observed in
the LASCO C2 field--of--view (FOV) at $\approx$ 13:30 UT. 
The speed and the angular width of this CME was 208  km s$^{-1}$ and 160$^{\circ}$
respectively.

\begin{figure}
\hspace*{-0.5cm}
\includegraphics[width=0.70\textwidth, angle=90, clip=]{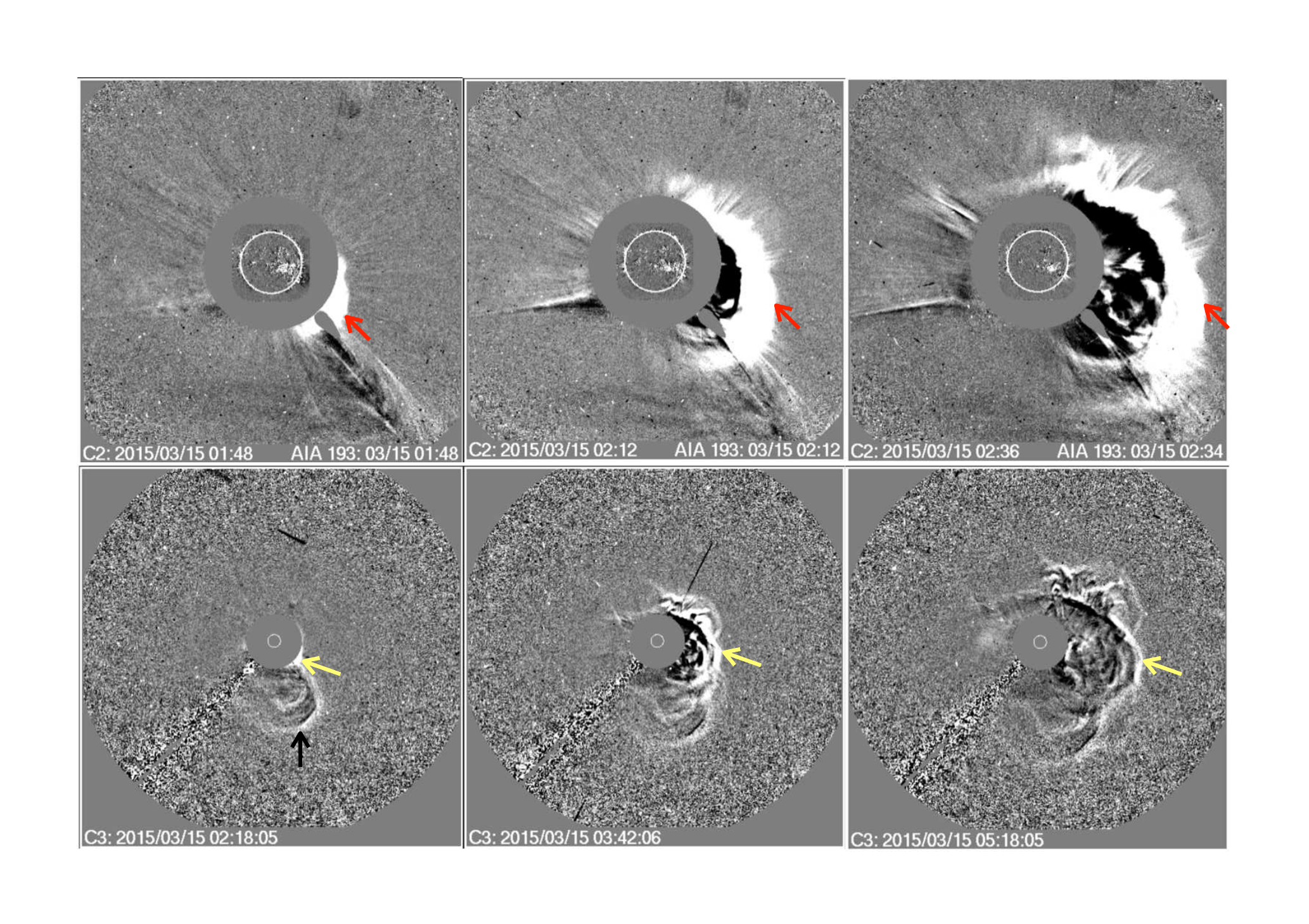}
%\vspace*{-1cm}
\caption{Development of CME on 15 March 2015 observed by LASCO C2 (top panel, red arrows) and C3 (bottom panel, yellow arrows). 
The black arrow indicates the CME of 14 March 2015.}   
\label{fig:cme}
\end{figure}

The F$_1$ filament eruption on 15 March 2015 produced a halo CME, visible in the LASCO C2 FOV at 01:48 UT. 
The CME was visible up to 27 R$_{\odot}$ in the LASCO C3 FOV.
The running difference of the C2 and C3 coronagraph images are shown in Figure \ref{fig:cme}.
The white circle represents the solar disk occulted by the coronagraph. In the C2 images, 
the running difference of SDO/AIA 193 \AA\ images are displayed inside the white circles for the same time. 
The red and yellow arrows indicate the leading 
edge of the CME in the LASCO C2 and C3 FOV respectively. The black arrow in the first image of the 
bottom panel points to the CME 
from the same active region on 14 March 2015. According to the LASCO Coordinated Data Analysis Workshop (CDAW) 
Catalog \citep{Gopalswamy09} the average CME speed 
was 719 km s$^{-1}$ and the acceleration was $-9.0$ m s$^{-2}$. 
The mass and kinetic energy of the CME were 3.0$\times 10^{16}$ gram and 7.7 $\times 10^{31}$ ergs respectively.	
Moreover, \inlinecite{Liu15} reported the maximum CME speed was 1100 kms$^{-1}$.
In the C3 FOV the CMEs of 14 and 15 March 2015  interacted at 02:18 UT. \inlinecite{Liu15} considered the interaction
of these CMEs as the cause of the strong geomagnetic storm of Solar Cycle 24 on 17 March 2015.

\section{Decay Index Calculations}
     \label{decay}

In order to find the reasons why the part F$_1$ of the initial filament erupted after a 12 hr delay in an 
intermediate state, while the remainder of the filament did not leave its position despite the 
strong activation, we should analyze the structure of the magnetic field surrounding these parts of the 
filament. Since the stability of the flux-rope equilibrium depends on the value of the decay index, we need 
to know the distribution of this parameter in the active region. In principle, the decay index should be 
calculated for the coronal field external to the flux rope. We can consider the electric current of the flux 
rope as the major coronal current within the volume of interest, while the surrounding magnetic field is 
created by currents located below the photosphere. For this part of the coronal field, which is needed for 
decay index calculations, we can use a potential-field extrapolation of the photospheric field data.

We are interesting in the  distribution of the potential magnetic field in the volume around the filament, which 
is much less than the size of the solar sphere. The bottom boundary of this domain is a part of the 
photosphere only slightly deviating from the shape of a flat surface. We cut a rectangular area of the 
photospheric magnetogram below the filament and use it as the boundary condition for the solution of the 
boundary-value problem in terms of Green's functions \citep{Filippov01,Filippov13}. 
Doing this we ignore the influence of sources that are located outside the chosen 
area. In many cases it can be reasonable.

In our case the active region NOAA AR 12297 is the strongest magnetic source on the disk and is rather distant 
from other active regions. On March 14 and especially on March 15 the active region is at a considerable 
distance from the central meridian, so we need to take into account the projection effect.

Figure 5a represents the modified fragment of the magnetogram taken by the HMI on 14 March 2015 
at 15:00 UT, which 
is used as a boundary condition for the potential magnetic field calculations. The pixel size in HMI 
magnetograms is  0.5$\arcsec$ or  0.36 Mm, which is very small compared with the expected height of a 
filament $\textgreater$ 10 Mm. 
To save computational time we apply binning several times and increase the pixel size to $\approx$ 
10 Mm. Figure 5b--f shows the distribution of the decay index 

\begin{equation}
n=-\frac{\partial \ln B_\textnormal t}{\partial \ln h}
\end{equation}

\noindent where {\it B}$_\textnormal t$ is the horizontal magnetic-field component and {\it h} is the height above the photosphere, at 
different heights above the area shown in the panel (a). The thin lines show isocontours of 
$\it n$ = 0.5, 1, 1.5, while the thick red lines indicate the positions of PILs at respective heights. 
Areas where $\it n$ $\textgreater$ 1 are tinted with grey, while regions with {\it n} $\textless$ 1 are white.
Green contours show the position of the filament taken from the 
co-aligned Kanzelhoehe H$\alpha$ filtergram transformed in the same way as the magnetogram 
(Figure~\ref{fig:decay_hal1}). 
Note that all panels (b)--(f) represent the 2D plots viewed from the normal-to-the-surface 
direction. Since the filaments are located at some unknown heights (we will consider this 
problem below) above this surface and the surface is inclined to the line-of-sight, the 
position of the filament contours does not correspond exactly to the position of magnetic 
features at any height (the filaments should be somewhat shifted to the north-west in 
this projection). However, the filament contours show approximately the places of our interest. 

A PIL is a favorable place for horizontal equilibrium of a flux rope, because the vertical 
component of the coronal field vanishes. Any PIL at any height can be considered as a 
potential location of the flux rope, but it is only a necessary condition. Another necessary 
(but again not sufficient) condition for stable equilibrium is that the decay 
index is below the critical value. In fact, flux ropes may be found only in few places where both 
conditions are fulfilled. In Figure~\ref{fig:decay}, the segments of PILs within white areas (or at least 
outside of isocontours 1.5) are the places favorable for the occurrence of a stable flux rope. 
Below 60 Mm (Figure~\ref{fig:decay}c) the segments of the PIL near both green contours are suitable for 
stable flux ropes. The contour $\it n$ = 1 touches the PIL near both filaments at the height of 75 Mm 
(Figure~\ref{fig:decay}d), while the contour $\it n$ = 1.5 touches the PIL near 
the southern filament at the 
height of 100 Mm (Figure~\ref{fig:decay}e) and near the western 
filament at the height of 120 Mm (Figure~\ref{fig:decay}f).

Unfortunately, we cannot measure directly the height of the filaments because STEREO is in an 
unfavorable position. However, we can estimate the heights using the method proposed by 
\inlinecite{Filippov16a}. 
It is based on the (confirmed by observations) assumption that the material of filaments 
is accumulated near coronal magnetic neutral surfaces B$_r$ = 0 
\citep{Filippov16b}.
It was found also that the potential 
approximation for coronal magnetic fields is sufficient for the filament height estimations. 
Comparison of the 3D shape of the neutral surface, represented in the projection on the 
plane of the sky as a set of PILs, with the filament shape and position allows us to obtain 
information about the heights of different parts of the filament including its top, or spine, which 
is the most reliable indicator of the flux-rope axis.    

Figure~\ref{fig:decay_hal1} shows the fragment of the Kanzelhoehe H$\alpha$ filtergram of the same region as 
in Figure~\ref{fig:decay} co-aligned with the magnetogram and transformed in the same way. 
The same PILs as in Figure~\ref{fig:decay}b--f are shown but every PIL is shifted in $\it x$ 
and $\it y$ coordinates by values

\begin{equation} 
\Delta x=htg\lambda_0
\end{equation} 

and

\begin{equation} 
\Delta y=htg\varphi_0 , 
\end{equation}

\noindent where $\it h$ is the height of the  PIL, $\lambda_0$ and $\varphi_0$ are the longitude and latitude of the selected area center. 
Thus they are projected on the plane of the sky in the same way as the filament in the on-disk 
filtergram. The lowest PIL at the height of 6 Mm is red, while the other is blue. The spine 
of the western filament follows exactly the PIL at the height of 30 Mm. All of the filament body is 
located between this line and the red line at the height of 6 Mm. The southern filament does 
not so strictly follow any PIL, however, its spine is most likely a little bit above the  PIL 
at the height of 78 Mm. For comparison Figure 7 shows the filament and the neutral surface 
on 14 March at 11 UT before the separation into two parts. The spine of the western section 
of the filament also follows exactly the PIL at the height of 30 Mm, while the eastern section 
seems to be higher. The top of the wide part of the eastern section touches the PIL at the height 
of 54 Mm and the thin thread-like continuation of the spine crosses all PILs.

Returning to Figure \ref{fig:decay} we establish that the western section of the filament before the 
separation and the western filament after the separation are relatively low (30 Mm) and are 
located in the zone of stability within the coronal magnetic field. The eastern section of 
the filament is less stable because on the one hand it is higher and on the other hand the 
decay index in this area is also higher. That is why, possibly, the disturbance coming 
from inner parts of the active region leads to the partial and failed eruption of the eastern 
section of the filament, which results in the separation of the filament into two parts. 
The erupted eastern section of the filament obviously finds a new equilibrium position at a 
greater height. In the new position of the eastern section, this height (about 80 Mm), now 
considered as the southern filament F$_1$, is within the zone of stability for the decay index 
threshold $\it n$$_c$ = 1.5 and on the edge of stability for the decay index threshold $\it n$$_c$ = 1. 
The next disturbance from the active region easily causes the start of the eruption of the 
southern filament and this eruption is full because there is no a zone of stability at heights 
above 100 Mm. So, the eastern section of the filament shows a two--step eruption with a 
metastable state at a height of 80 Mm for 12 hrs. Possibly it could stay there longer if the 
strong disturbance did not come from inner parts of the active region. The western filament is 
deep within the zone of stability and therefore it did not erupt despite the strong activation by 
the energetic disturbance.

\section{Discussion and Conclusion}
     \label{dicussion}

In this paper, we have presented the two--step filament eruption of 14--15 March 2015 using 
observations from AIA onboard the SDO satellite
and ground--based GONG and {\em Kanzelhoehe Solar Observatory} H$\alpha$ data. Our main results are as follows:
\\
\begin{itemize}

\item
The initiation of the filament eruption on 14 March 2015 and its full eruption
 on 15 March 2015 are associated with the jets
in the active region.
\\

\item
The decay index calculation suggests that on 14 March 2015 the filament first enters into the instability zone and after
reaching some height it finds itself in the stability zone. Again on 15 March 2015 the filament enters into the instability
zone and finally it erupts.
\\

\item
The major part of filament which had not been destroyed on 14 March 2015 was activated on March 15 but could not erupt. Therefore
it was a failed eruption. The coronal magnetic field calculation evidences that the  decay index at the filament location is below the 
threshold of the torus instability and hence the filament fails to erupt.
\\

\item
The observation of the same sign of the twist/helicity in the chromosphere, higher 
solar atmosphere and in the magnetic cloud evidence the
conservation property of the helicity.
\\
\end{itemize}

We find that a big filament located at the periphery of a strong active region undergoes
a complicated partial and two-step eruption. The idea of two-step energy release processes 
came from analyses of two-peak EUV light-curves of some flares \citep{Woods11,Su12} suggesting that 
two peaks in light-curves appear due to two 
stages of a single event associated with the delayed eruption of a CME. \inlinecite{Woods11} and \inlinecite{Su12}
 presented AIA EUV observations of the limb 8 March 2011 event in which a flux rope accelerated in the first 
stage up to 120 km s$^{-1}$, then the speed decreased to 14 km s$^{-1}$, and in the second stage, started 
after 2 hrs after the beginning of the event, it accelerated again and became the CME with 
a speed of $\approx$ 500 km s$^{-1}$. \inlinecite{Byrne14} also analyzed this event and suggested that either 
the kink-instability or the torus-instability of the flux rope was the likeliest scenario. 
Since the event was at the limb and photospheric magnetic-field data were not available for 
this time, the authors did not make strong conclusions about the magnetic configuration and were 
not very certain about the supposed torus instability without calculation of the decay index. 
\inlinecite{Gosain16} studied a two-step eruption of a quiescent filament on 22 October 2011. 
It was observed from different viewpoints by SDO, SOHO, and STEREO. The CME associated with 
the filament eruption and two bright ribbons in the chromosphere both appeared 15 hrs after 
the start of the event. Computation of the decay index showed that there were zones of 
stability and instability that alternated in the corona. Below 100 Mm the equilibrium was 
stable, then the zone of instability followed from 100 to 500 Mm and gave place to a 
zone of stability again. Above a height of 600 Mm the  PIL disappeared, which hinted at the 
possibility for the flux rope to lose the horizontal equilibrium and erupt. These results 
showed the possible scenario of the two-step eruption confirmed by observations and 
calculations. However, the flux rope was not clearly observed in the intermediate 
position and the magnetic field calculations at great heights, above 400 Mm, were not too reliable. 

The event on 14--15 March 2015 we  analyze in this paper happens when the initial filament is in 
positon favorable for observations somewhere between the central meridian and the limb. This 
allowes us to have reliable photospheric magnetic-field data and to observe the ascending 
trajectory of the eruptive filament. The disadvantage is the absence of STEREO observations 
from another point of view, but we use the rather reliable method of measuring of the filament 
spine height on the disk. Our calculations of the decay index of the coronal magnetic field 
show that after the first loss of equilibrium a part of the filament finds itself within 
the zone of stability, but not far from the threshold of instability. The strong disturbance 
that comes from the inner parts of the active region force the filament to enter into the second 
step of the eruption. 

\begin{figure}
\includegraphics[width=1.0\textwidth,clip=]{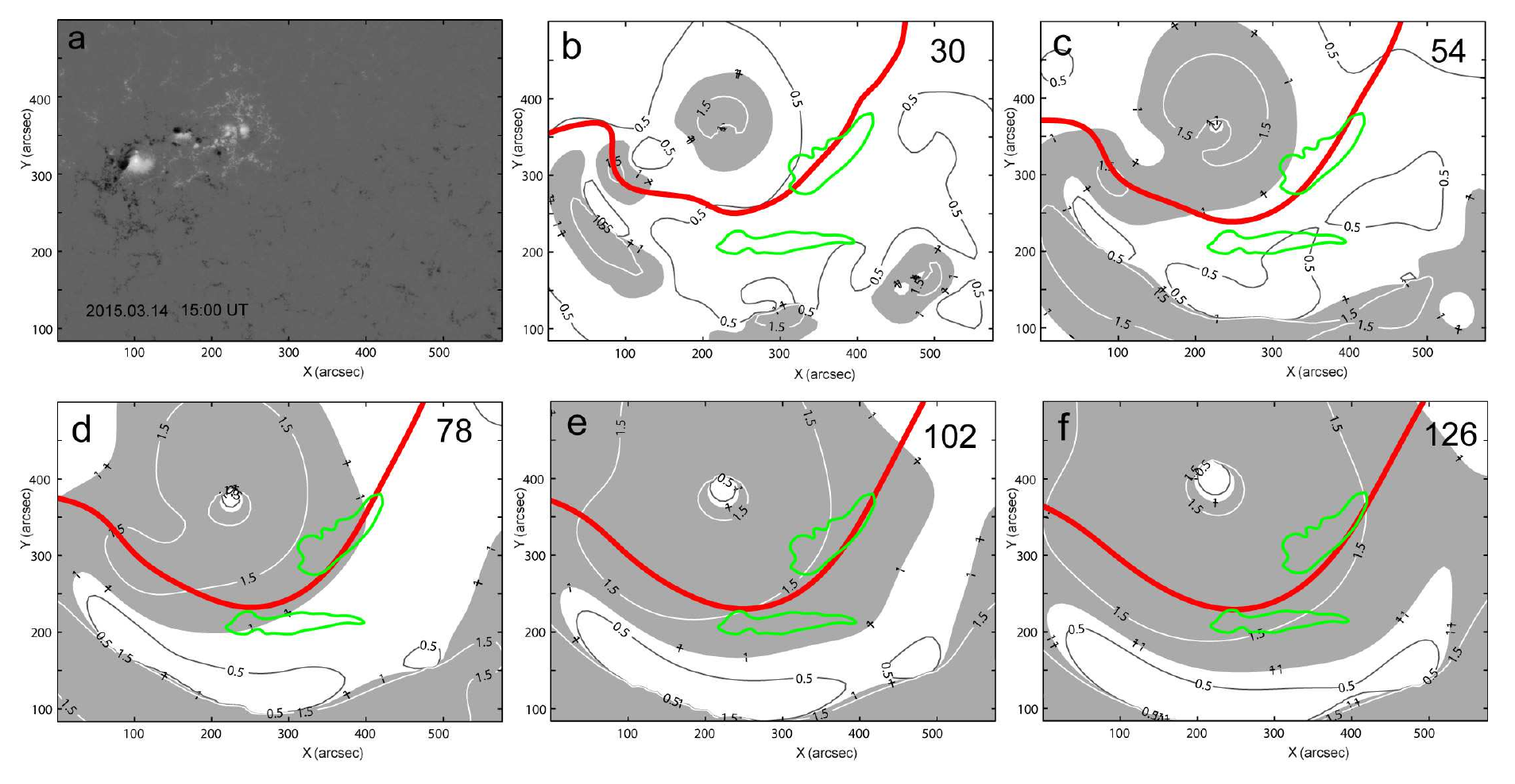}
\caption{(a) HMI magnetogram of the selected area taken on 14 March 2015 at 15:00 UT. Panels (b)-–(f) distribution of 
the decay index $\it n$ at different heights in Mm above the area shown in 
panel (a). A red line indicates the position of the PILs. 
Areas where $\it n$ $\textgreater$ 1 are tinted with grey, while regions with {\it n} $\textless$ 1 are white.
Green contours show the position of the filament taken from the co-aligned Kanzelhoehe H$\alpha$ filtergram.
}
\label{fig:decay}
\end{figure}

\begin{figure}
\includegraphics[width=1.0\textwidth,clip=]{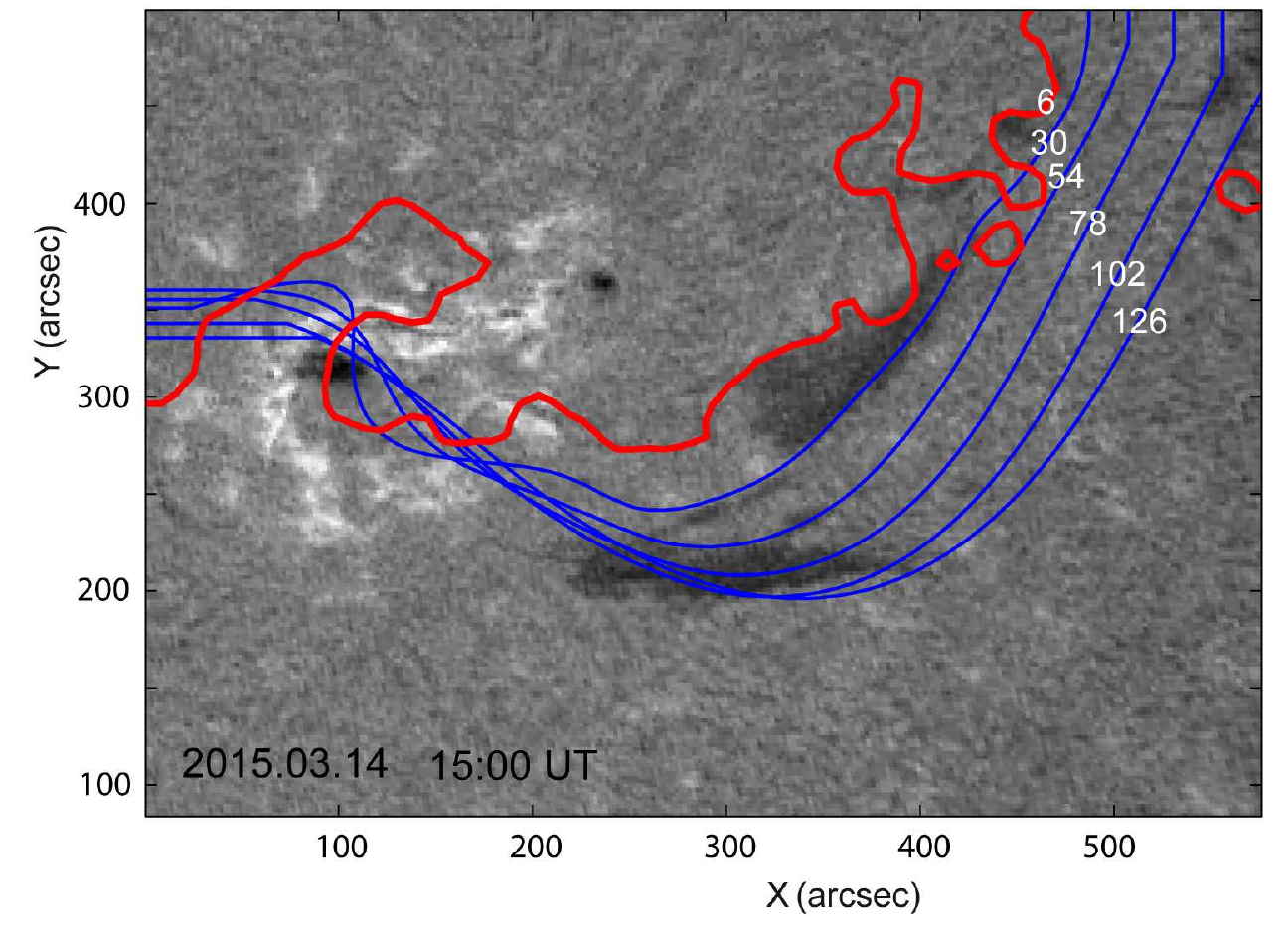}
\caption{Kanzelhoehe H$\alpha$ filtergram of the same region as in 
Figure 5a taken at the same time with superposed PILs at different heights.} 
\label{fig:decay_hal1}
\end{figure}

\inlinecite{Wang16} studied the filament eruption on 15 March 2015 as the source of the strong 
geomagnetic storm of Solar Cycle 24. They analyzed the second step of the eruption because 
it was this event that produced the fast CME that together with another CME started 12 hrs 
earlier drove the geomagnetic storm. 
They calculated the decay index $\it n$ in the region using the potential-field source-surface (PFSS) model. They found 
that {\it n} reached the critical value 1.5 for the onset of the torus instability at a height of 
approximately 118 Mm and concluded that the strapping field was strong enough to prevent any 
eruption of both the southern and western filaments. If their decay index calculation is averaged over the 
whole active region it is similar to our result shown in Figure \ref{fig:decay}f. 
They related  the abrupt acceleration 
of the southern filament around 01:15 UT to the passage of the filament into the region 
with open magnetic field, which did not further confine the filament. However, it is difficult to 
derive the true location of the filament rather high in the corona from images in only one projection.

In our study, we perform the analysis of the  filament first-step activation on March 14 
and the second step eruption on March 15. We explain this as follows, propogating a 
 scenario based on more accurate magnetic field calculations. Our conclusions 
are based on the detailed distribution of the decay index over the active region. We define 
the critical height from values of the decay index in the places where the filament could 
be located and is really located. The distribution of the decay index provides an 
opportunity for the two-step eruption. The estimated heights of the filaments show 
that the western filament is within the zone of stability during all times of 
observation. That is why it keeps its position despite the strong disturbance and 
intensive internal motions. The southern filament is close to the threshold of 
stability during all time before the abrupt acceleration in the second step. Very 
likely the eastern part of the initial filament becomes slightly unstable at the 
beginning of the first step. It moves rather slowly in the direction of the zone 
of stability and is able to find a new position for stable equilibrium there. 
However, it is also close to the threshold of stability and the next disturbance 
from the active region causes the full eruption.

%%%%%%%%%%%%%%%%%%%%%%%%%%%%%%%%%%%%%%%%%%%%%%%%%%%%%%%%%%%%%%%%%%%%%%%%%%%%%%%%%%%%%
%%%%%%%%%%%%%%%%%%%%%%%%%%%%%%%%%%%%%%%%%%%%%%%%%%%%%%%%%%%%%%%%%%%%%%%%%%%%%%%%%%%%%

%%%%%%%%%%%%%%%%%%%%%%%%%%%%%%%%%%%%%%%%%%%%%%%%%%%%%%%%%%%%%%%%%%%%%%%%%%%%%%%%%%%%%
%%%%%%%%%%%%%%%%%%%%%%%%%%%%%%%%%%%%%%%%%%%%%%%%%%%%%%%%%%%%%%%%%%%%%%%%%%%%%%%%%%%%%

%%%%%%%%%%%%%%%%%%%%%%%%%%%%%%%%%%%%%%%%%%%%%%%%%%%%%%%%%%%%%%%%%%%%%%%%%%%%%%%%%%%%%
%%%%%%%%%%%%%%%%%%%%%%%%%%%%%%%%%%%%%%%%%%%%%%%%%%%%%%%%%%%%%%%%%%%%%%%%%%%%%%%%%%%%%

%%%%%%%%%%%%%%%%%%%%%%%%%%%%%%%%%%%%%%%%%%%%%%%%%%%%%%%%%%%%%%%%%%%%%%%%%%%%%%%%%%%%%
%%%%%%%%%%%%%%%%%%%%%%%%%%%%%%%%%%%%%%%%%%%%%%%%%%%%%%%%%%%%%%%%%%%%%%%%%%%%%%%%%%%%%

%%%%%%%%%%%%%%%%%%%%%%%%%%%%%%%%%%%%%%%%%%%%%%%%%%%%%%%%%%%%%%%%%%%%%%%%%%%%%%%%%%%%%

%%%%%%%%%%%%%%%%%%%%%%%%%%%%%%%%%%%%%%%%%%%%%%%%%%%%%%%%%%%%%%%%%%%%%%%%%%%%%%%%%%%%%
%%%%%%%%%%%%%%%%%%%%%%%%%%%%%%%%%%%%%%%%%%%%%%%%%%%%%%%%%%%%%%%%%%%%%%%%%%%%%%%%%%%%%

%%%%%%%%%%%%%%%%%%%%%%%%%%%%%%%%%%%%%%%%%%%%%%%%%%%%%%%%%%%%%%%%%%%%%%%%%%%%%%%%%%%%%
%%%%%%%%%%%%%%%%%%%%%%%%%%%%%%%%%%%%%%%%%%%%%%%%%%%%%%%%%%%%%%%%%%%%%%%%%%%%%%%%%%%%%

%%%%%%%%%%%%%%%%%%%%%%%%%%%%%%%%%%%%%%%%%%%%%%%%%%%%%%%%%%%%%%%%%%%%%%%%%%%%%%%%%%%%%
\begin{figure}
\includegraphics[width=1.0\textwidth,clip=]{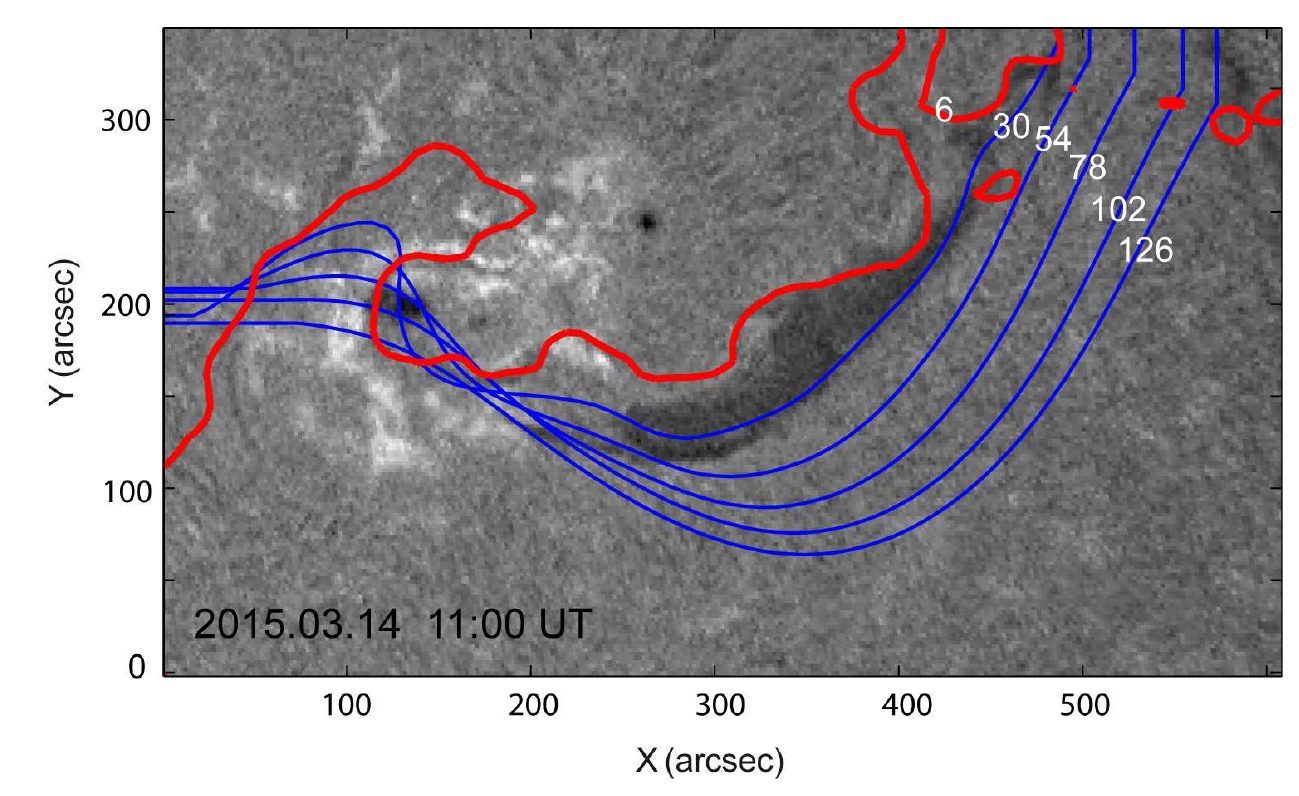}
\caption{Kanzelhoehe H$\alpha$ filtergram on 14 March 2015 at 11:00 UT with the filament before the 
separation with superposed PILs at different heights.}
\label{fig:decay_hal2}
\end{figure}
%%%%%%%%%%%%%%%%%%%%%%%%%%%%%%%%%%%%%%%%%%%%%%%%%%%%%%%%%%%%%%%%%%%%%%%%%%%%%%%%%%%%%

\begin{acks}
We thank the anonymous referee for the valuable comments and suggestions.
We acknowledge the use of SDO and GONG, and the Kanzelhoehe Solar Observatory data. RC acknowledges the support from
SERB--DST project no. SERB/F/7455/ 2017-17. RJ thanks the Department of Science and Technology (DST),
Government of India for an INSPIRE fellowship. BS wants to thank the team of the ISSI workshop managed by Nicolas Labrosse 
on ``Nature of prominences'' in Bern for fruitful discussion on this event.

\noindent {\bf Disclosure of Potential Conflicts of Interest}  The authors declare that they have no conflicts of interest.
\end{acks}
%%%%%%%%%%%%%%%%%%%%%%%%%%%%%%%%%%%%%%%%%%%%%%%%%%%%%%%%%%%%%%%%%%%%%%%%%%%%%%%%%%%%%%%%%%%%%%%%%%%%%%%%%%%%%%

\mbox{}~\\
\bibliographystyle{spr-mp-sola}
\bibliography{reference_new}
\IfFileExists{\jobname.bbl}{} {\typeout{}
\typeout{***************************************************************}
\typeout{***************************************************************}
\typeout{** Please run "bibtex \jobname" to obtain the bibliography}
\typeout{** and re-run "latex \jobname" twice to fix references}
\typeout{***************************************************************}
\typeout{***************************************************************}
\typeout{}}

\end{article}
\end{document}